\documentstyle[11pt,fleqn]{article}
\newcommand{\ret}{\nonumber \\}
\newcommand{\beq}{\begin{equation}}
\newcommand{\eeq}{\end{equation}}
\newcommand{\beqar}{\begin{eqnarray}}
\newcommand{\eeqar}{\end{eqnarray}}

\newcommand{\eps}{\epsilon}
\newcommand{\G}{\Gamma}

\newcommand{\vareps}{\varepsilon}
\newcommand{\al}{\alpha}
\newcommand{\bet}{\beta}
\newcommand{\df}{\partial}
\newcommand{\om}{\omega}
\newcommand{\sig}{\sigma}
\newcommand{\hpsi}{{\widehat \psi}}
\newcommand{\hPsi}{{\widehat \Psi}}
\newcommand{\bin}[2]{C^{{#1}}_{{#2}}}

\font\titlefnt=cmbx10 scaled \magstep2

\begin{document}
\mathindent 0mm
\setcounter{page}{0}
\newpage\thispagestyle{empty}
\begin{flushright}
SPbU--IP--96--30
\end{flushright}
\topskip 2cm
\vspace*{1cm}
\begin{center}
\boldmath
{\titlefnt High--gradient operators in the $N$--vector model\\
\vspace*{0.4cm}
}
\unboldmath
\vskip3cm
S.E.~Derkachov\footnote[1]
{E--mail: manashov@phim.niif.spb.su}
 \\
Department of Mathematics, St.~Petersburg Technology Institute\\~\\
S.K.~Kehrein\footnote[2]
{E--mail: kehrein@physik.uni-augsburg.de\\
New address: Institut f\"ur Physik der Universit\"at
Augsburg, Theoretische Physik III, Elektronische Korrelationen
und Magnetismus, D--86135 Augsburg, Germany} \\
Institut f\"ur Theoretische Physik,
Ruprecht--Karls--Universit\"at Heidelberg, \\
Philosophenweg~19, D--69120 Heidelberg, Germany \\~\\
A.N.~Manashov${}^{1}$ \\
Department of Theoretical Physics, State University St.~Petersburg\\
\vspace*{1.5cm}
\today
\\
\vspace*{2cm}
{\bf Abstract}
\\

\end{center}
\vspace*{1cm}
It has been shown by several authors that a certain
class of composite operators with many fields and gradients endangers
the stability of nontrivial fixed points in $2+\eps$~expansions for
various models. This problem is so far unresolved. We
investigate it in the $N$--vector model in an $1/N$--expansion.
By establishing an asymptotic naive addition law for anomalous
dimensions we demonstrate that the first
orders in the $2+\eps$~expansion can lead to erroneous
interpretations for high--gradient operators. While this makes
us cautious against over--interpreting such expansions (either
$2+\eps$ or $1/N$), the stability problem in the $N$--vector model
persists also in first order in $1/N$ below three dimensions.

\vfill
\pagebreak
\topskip 0cm

\renewcommand{\theequation}{\thesection.\arabic{equation}}
\setcounter{equation}{0}
\section{Introduction}
Problems with high--gradient operators were first observed by
Kravtsov, Lerner and Yudson in the $Q$--matrix model in a
$2+\eps$~expansion~\cite{KLY1, KLY2}. They noticed that a certain class
of canonically irrelevant composite operators with $2s$~gradients
acquires anomalous dimensions in first order in~$\eps$ that grow
with~$s^2$ and that finally make an infinite number of these operators
relevant for $s\rightarrow\infty$ and any fixed finite~$\eps>0$.
Wegner et al. have shown that this behaviour occurs also in
other $2+\eps$~expansions in one--loop order: In the $N$--vector model
\cite{W1}, the unitary matrix model \cite{W2} and the orthogonal
matrix model \cite{MW}. In fact this behaviour has been
observed in all $2+\eps$~expansions that have so far been investigated
with respect to this question. Judging from these results one could argue
that the conventional nontrivial fixed points of these models
become unstable against an infinite number of high--gradient perturbations.
This would have important consequences for the present understanding
of many problems that rely on $2+\eps$~expansions.

A natural starting point to investigate the role of high--gradient
operators in more detail is the universality class of the
$N$--vector model. Here various expansions are available:
$2+\eps$, $1/N$ and $4-\eps$~expansions. In Ref.~\cite{W1}
Wegner found a composite operator with $2s$~gradients
and the full scaling dimension
\beq \label{hg1}
y=2(1-s)+\eps \left( 1+\frac{s(s-1)}{N-2} \right) +O(\eps^2)
\eeq
in the $2+\eps$~expansion of the nonlinear $\sigma$--model
\beq
S=\frac{1}{2T}\int d^dx\:(\partial_\mu \vec\pi)^2 \ .
\label{sigmamodel}
\eeq
Here
$\vec\pi=(\pi_1,\ldots,\pi_N)$ is the normalized field
$\vec\pi^2=1$ of the nonlinear $\sigma$--model.
Operators with $y>0$, $y=0$ and $y<0$ are relevant, marginal and
irrelevant, resp. Castilla and Chakravarty have extended the
calculation to two--loop order \cite{CC1,CC2} and obtained the term in
order~$\eps^2$
\beq \label{hg2}
y=2(1-s)+\eps \left( 1+\frac{s(s-1)}{N-2} \right)
+ \frac{\eps^2 s^3}{(N-2)^2} \left(\frac{2}{3}+O(1/s)\right)+O(\eps^3) .
\eeq
This two--loop result makes the stability problem even worse.
For $s\gg N/\eps$ Eq.~(\ref{hg2}) amounts to an infinite number of
relevant operators. On the other hand it could
be proven in Ref.~\cite{KWP} that no similar problem
occurs in the $4-\eps$~expansion of $(\vec\phi^2)^2$--theory.
In the $4-\eps$~expansion all anomalous dimensions in order~$\eps$
can only make composite operators more irrelevant, but never
go in the ``dangerous" relevant direction.

Therefore somehow the behaviour seems to bend over between
$d=2$ and $d=4$ dimensions. The obvious tool to investigate
this question is the $1/N$--expansion for $2<d<4$. This is
the programme of this paper. Along its lines two other
interesting aspects of the $N$--vector model emerge.
First of all it is shown how
the complete spectrum of anomalous dimensions for a
subalgebra of composite operators in order~$1/N$ is
encoded as a straightforward diagonalization problem.
This formally solves part of a programme by
Lang and R\"uhl \cite{LR}.
Then an asymptotic naive addition law for anomalous
dimensions is derived that has also been found in the
$4-\eps$~expansion \cite{K,DM}. This leads to
a ``clustering" of anomalous dimensions.
However, this property is lost in a $2+\eps$ expansion.
It will become apparent that this serves as a warning
against using the first orders of the $2+\eps$~expansion
to make statements about the high--gradient
limit~$s\rightarrow\infty$.

Still unfortunately this is not the end of the story. From an
older result by Vasil'ev and Stepanenko \cite{VS}
we will see that the high--gradient problem persists
for $2<d<3$~dimensions also after adding up all order in~$\eps$
in first order in~$1/N$. From their work one can see that
powers~$\sigma^s$ of the auxiliary field~$\sigma$ in the
$1/N$--expansion interpolate between the high--gradient
operators in $2+\eps$ and composite operators~$(\vec\phi^2)^s$
in the $4-\eps$~expansion. For some reason this simple fact has
so far remained unnoticed and we also want to use this paper to
clarify this point.

Our paper is organized as follows. In Sect.~2 we discuss the
renormalization of a certain class of composite operators
to first order in the~$1/N$--expansion. Eigenoperators and
their critical exponents are encoded as a
straightforward diagonalization problem similar to
Ref.~\cite{KWP}. In Sect.~3 an asymptotic naive addition
law of anomalous dimensions valid for a large number
of gradients is found that has already been
observed in the $4-\eps$~expansion \cite{K,DM}. However,
it is shown that this structure remains unobserved in the
$2+\eps$~expansion which can lead to misleading
interpretations. In Sect.~4 it is shown that results in first
order in~$1/N$ obtained by Vasil'ev and Stepanenko in
Ref.~\cite{VS} allow the interpolation of the high--gradient
behaviour between two and four dimensions. Sect.~5 contains the
conclusions and suggestions for future work to solve
the stability problem.

\section{The spectrum of anomalous dimensions in order~$1/N$}
\setcounter{equation}{0}

\subsection{The operator subalgebra ${\cal C}_{\rm sym}$}
\label{conformal}

In this paper we shall mainly deal with composite operators
$\Psi_{i_1,i_2...i_l}^{A_1,A_2...A_n}(x)$
constructed from n fields $\phi_A$ ($A = 1,...,N$) and
their derivatives which are symmetric and traceless both
with respect to their spatial $O(d)$ and internal $O(N)$
indices. We denote this important class of composite operators
by~${\cal C}_{\rm sym}$, compare also Ref.~\cite{KW}.

It is convenient to introduce ``scalar" composite operators
\begin{equation}
\Psi(x,u,t) \equiv
\Psi_{i_1...i_l}^{A_1...A_n}(x)
u^{i_1}\ldots u^{i_l}t_{A_1}\ldots t_{A_n},
\label{oper1}
\end{equation}
where $u$ and $t$ are $d$--dimensional  and
$N$--dimensional vectors, resp.
The original composite operators
$\Psi_{i_1,i_2...i_l}^{A_1,A_2...A_n}(x)$
can be reconstructed from $\Psi(x,u,t) $ by differentiation
with respect to $ u $ and $ t $.

Now $\Psi(x,u,t) $ can generally be expressed as
\begin{equation}
\Psi(x,u,t) =
\psi(\frac{\df}{\df a_1},...,
\frac{\df}{\df a_n})
\Phi(x;u;t;a_1,...,a_n) \biggl|_{a_{i} = 0},
\label{def1}
\end{equation}
where
\begin{equation}
\Phi(x;u;t;a_1,...,a_n) \equiv
 P(u,\df_{v})
T(t,\df_{s})
\prod_{m=1}^n s^{A_{m}}\phi_{A_{m}}(x+a_m v)\biggl|_{v,s=0}.
\label{def2}
\end{equation}
Here $ P $ and $ T $  are projectors on the subspace of
traceless and symmetric tensors. The action of
$ P $ (resp. $ T $) on the operator $\Psi(x,u,t)$
amounts to the elimination of all terms proportional to
$ u^{2} (resp. t^{2}) $. The explicit form of the projector
$ P,T $ can be found in Ref.~\cite{DP}, but will not be used in the
following.

In this manner Eq.~(\ref{def1}) establishes a one to one
correspondence between a composite operator
$\Psi(x,u,t)\in {\cal C}_{\rm sym} $ and a symmetric
polynomial
$ \psi(z_{1},\ldots,z_{n}) $, which we call the coefficient
function of the operator $\Psi(x,u,t)$.

Now we briefly discuss the important subclass of conformal operators in
${\cal C}_{\rm sym} $ that will
play a central role in our further investigations.
The operator
$\Psi_{i_1,i_2...i_l}(x)$ (for a moment we omit the internal indices)
is called conformal if it transforms
as a tensor under conformal coordinate transformations, i.e.
\begin{equation}
\delta_{\alpha}\Psi_{i_1,\ldots,i_l}(x) =
(\alpha\df + \frac{\Delta_{\Psi}}{d}
\df \alpha )\Psi_{i_1,\ldots,i_l}(x) +
\sum_{k=1}^l \df_{i_k} \alpha_s
\Psi_{i_1,..,s,..,i_l}(x)
\label{def_conf}
\end{equation}
at
$$
x \rightarrow x^{\prime}(x)= x+\alpha(x),\>\>\>
\delta_{\alpha}\phi(x) \equiv
(\alpha(x)\df + \frac{\Delta}{d}\df \alpha (x))\phi(x),
$$
with
$
\alpha_i(x) = a_i + \omega_{ik}x_k + \lambda x_i +
(x^2\delta_{ik} - 2x_ix_k)c_k.
$ \\
Here
$ \Delta $ and
$ \Delta_{\Psi} $ are the conformal dimensions of the elementary field
$ \phi(x) $ and the operator $\Psi_{i_1,\ldots,i_l}(x)$, resp.
Note that for conformal operators from ${\cal C}_{\rm sym} $
one simply has $\Delta_\psi$ =
({\rm scale dimension}) - ({\rm number of spatial indices}).)

A straightforward calculation shows that the coefficient function
$ \psi $ of conformal operators must satisfy the following equation
\begin{equation}
\sum_{m=1}^n (z_m \df^{2}_{z_m} +
\Delta\df_{z_m}) \psi(z_1,\ldots,z_n) = 0\ .
\label{confeq}
\end{equation}
To find all solutions of this equation let us define the one
to one map
$ \psi\to\hpsi $:
\begin{equation}
\hpsi(z_1,\cdots,z_n) =\mbox{\bf T}\psi(z_1,\cdots,z_n)=
\frac{1}{\Gamma^n(\Delta)}
\int_{0}^{\infty}
\prod_{m=1}^n {\rm d}t_m \ t_m^{\Delta - 1}{\rm e}^{-t_m}
\psi(t_1 z_1 ,...,t_n z_n) .
\label{hatpsi}
\end{equation}
For the function
$ \hpsi $ Eq.~(\ref{confeq}) reads
\begin{equation}
\label{transl}
\sum_{m=1}^n \df_{z_m}
\hpsi(z_1,\ldots,z_n) = 0,
\end{equation}
that is $ \hpsi $ must be translationally invariant.
Thus the space of conformal operators with
$ n $ fields and
$ l $ derivatives is isomorphic to the vector space
 $\hat C(n,l)$ of symmetric, homogeneous of degree~$l$
and translationally invariant polynomials of n variables.
The dimension of $\hat C(n,l)$
can be easily calculated from the generating function \rm{D}(n,x)
\cite{KW}
\begin{equation}
\label{dim}
\rm{D}(n,x) \equiv
\sum_{l=0}^{\infty} \rm{dim}C(n,l)\> x^l =
\frac{1}{\prod_{p=2}^{n}(1-x^p)}.
\end{equation}

\subsection{Calculation of Feynman diagrams}

To describe the renormalization of operators from
${\cal C}_{\rm sym}$ in the nonlinear $\sigma$--model we follow the
scheme developed in the papers~\cite{VS,VN}.
The regularized action for the theory under consideration has the form
\begin{equation}
S = -\frac{1}{2}\df_{\mu}\phi^{A}\df^{\mu} \phi_{A}
 -\frac{1}{2}\sig K_{\om} \sig
 +\frac{1}{2}M^{\om}\sig
\phi^{A}\phi_{A}
 +\frac{1}{2}M^{2\om}\sig K \sig \ .
\label{action}
\end{equation}
Here
$\om$
and
$ M $
are regularization parameter and
regularization mass, resp.
The kernel
$ K $
is determined from the requirement of the cancellation
of self--energy diagram contributions of the sigma field
(see Fig.~1);
$ K(x)=-N/2G_{\phi}^{2}(x) $,
where
$ G_{\phi}(x) $
is the propagator of
the $ \phi^{A} $
field
($<\phi^{A}(x)\phi^{B}(y)>=\delta^{AB}G_{\phi}(x)$)~\cite{VN}.

The regularized kernel
$ K_{\om} $
is defined as
$K_{\om}(x)=K(x)x^{-2\om}$.\\
The explicit expressions for the propagators
$ G_{\phi} $ and
$ G_{\sig}=K_{\om}^{-1} $ in momentum space are
\begin{equation}
 G_{\phi}(p)=p^{-2},\>\>\>\>\> G_{\sig}(p)=B p^{-2(\mu-2+\om)},
\end{equation}
where
$ \mu\equiv d/2 $  and
$ B=-{2\cdot 4^{\mu+\om}\pi^{\mu}}
{\Gamma(2\mu-2+\om)}/\left({N \Gamma^{2}(\mu-1)}{\Gamma(2-\mu-\om)}\right) $.

In the calculation of Feynman diagrams the divergencies appear as poles
in $ \om $ which therefore corresponds to
$ \epsilon $ in the dimensional regularization scheme.
It should be noted that the nonlinear $\sigma$--model is not multiplicatively
renormalizable~\cite{VN}. In general its critical exponents
cannot be expressed via the renormalization constants
$ Z_{i} $ but must be determined (employing conformal invariance of the
theory~\cite{VN}) directly from the renormalized Green functions by
looking at the long distance logarithmic contributions. However,
in Ref.~\cite{VS} it was
shown  that in order to find the critical exponents in first order in
$ 1/N $
it is sufficient to calculate the divergent pole term
of the Feynman diagrams. A detailed discussion of this point can
be found in~\cite{VS}.

Thus to find the critical exponents in first order in
$ 1/N $
we must extract the divergent part in the corresponding diagrams.

For the operators from
${\cal C}_{\rm sym}$
only the diagram in Fig.~2 contributes in first order in
$ 1/N $. Diagrams containing
cycles of
$ \phi $ lines are zero due to the
$ O(N) $ tracelessness of the operators in question. Diagrams
describing transitions
$ \df_{i}\df_{k}\to \delta_{ik}\sig $ vanish due to
the tracelessness in the spatial indices.

As usual one  has to consider the renormalization of the set of mixing
operators. Renormalized operators are defined as
\begin{equation}
[\Psi_{r} (\phi)]_R =
\Psi_{r}(\phi) + \frac{1}{\om}
\sum_{r'} Q^{rr'} \Psi_{r'} (\phi) \ .
\label{renO}
\end{equation}
The critical exponents are obtained from the mixing matrix
$ Q $ via
\begin{equation}
x_{\Psi}^{rr'}(g) =
n \gamma_{\phi} \delta^{rr'} +
 2 {Q^{rr'}},
\end{equation}
where $n$ is the number of fields $\phi$ in the operator $\Psi_r(\phi)$
and
$\gamma_{\phi}=\eta/2$ is the anomalous dimension of
the elementary field~$\phi$ with
\beq
\eta= \frac{4}{N}\,\frac{\G(2\mu-2)}{\G^2(\mu-1)\G(2-\mu)}\,
\frac{2-\mu}{\G(\mu+1)}
 +O(1/N^2) \ .
\label{adim_phi}
\eeq

It is convenient to fulfill the renormalization procedure in
terms of the generating function~(\ref{def2}). The analytic expression
corresponding to the diagram shown in Fig.~2 is
\beq
e^{\imath(u\, ,\, p_{i}a_{i}+p_{k}a_{k})}
\int \frac{d^{2\mu}l}{(2\pi)^{2\mu}}\:
\frac{e^{\imath(a_{i}-a_{k})(u\, ,\,l)}}
{l^{2(\mu-2+\om)}(l+p_{i})^{2}(p_{k}-l)^{2}} ,
\eeq
where $(\cdot,\cdot)$ denotes the scalar product.
The calculation of this integral does not cause any problems and we
give the final answer for the renormalization of the
generating function at once
\begin{equation}
\label{R0}
[\Phi(x;u;t;a_1,...,a_n)]_R  =  \Phi(x;u;t;a_1,...,a_n) +
\frac{\nu(\mu)}{\om N} \sum_{i<k} H_{ik}\Phi(x;u;t;a_1,...,a_n) \ ,
\end{equation}
where
\beq
\nu(\mu)=
2\:\frac{\Gamma(2\mu-2)}{\Gamma^{2}(\mu-1)\Gamma(2-\mu)\Gamma(\mu-2)}
\label{def_nu}
\eeq
and
\begin{eqnarray}
&&{H_{ik}\Phi(..a_i,..,a_k,..) =1/2(h_{ik}+{\bar h}_{ik})
\Phi(..a_i,..,a_k,..)=}\ret
\lefteqn{ 1/2 \biggl( \int^1_0
\prod_{m=1}^{3}{\rm d}\alpha_m
\delta (\sum_{n=1}^{3}\alpha_n - 1)
\alpha_2^{\mu - 3}\Phi(..,\alpha_1 a_i + {\bar \alpha_1}a_k,..,
\alpha_3 a_k+{\bar \alpha_3}a_i,..)+(i\leftrightarrow k)\biggr)}
\label{R1}
\end{eqnarray}
with the notation
$ {\bar \alpha_{m}}\equiv (1-\alpha_{m}) $.
On the rhs of this integral equation the $i$--th and
$k$--th arguments are replaced, the other arguments (denoted
by dots) remain unchanged. The first term in the brackets
in~Eq.~(\ref{R1}) determines the action of the operator
$ h_{ik} $ on the function
$\Phi(..,a_i,..,a_k,..)$, while the representation for ${\bar h}_{ik}$
is obtained from the latter by interchanging
$ a_{i}\leftrightarrow a_{k} $. Although on symmetric functions
$ h_{ik} $ and ${\bar h}_{ik}$  are identical, it will be useful
to distinguish them in the sequel.

To avoid misleadings we stress that~(\ref{R0}) should be
understood as an equation for generating functions, i.e. as an
equation for the coefficients in the power series expansion in
$ a_{i} $ of the lhs and rhs expressions. Eq.~(\ref{R1})
also requires some comments. Due to the factor
$\alpha_2^{\mu - 3} $  the integrand has a nonintegrable singularity and
the integral in rhs of Eq.~(\ref{R1})
must be understood as the analytical continuation from the region
$ \mu >2 $. As a matter of fact
$  \alpha_2^{\mu - 3} $ is always multiplied by some polynomial in
$ \alpha_{i} $ and the integral can always be expressed in terms of
$ \Gamma$--functions.

\subsection{Diagonalization problem}

To obtain the spectrum of critical exponents we must diagonalize
the mixing matrix
$ Q $, i.e. find the multiplicatively renormalized eigenoperators.
Taking into account Eqs.~(\ref{def1}), (\ref{renO}), (\ref{R0})
it is easy to derive the eigenvalue equation for
$ Q $. Indeed, using the Fourier representation for
\mbox{$ \phi(x+a_{i}u) $}
($ \phi(x)=\int d^{2\mu}q\,\phi(q){\rm e}^{{iqx}}$) one immediately
finds that the operator
$ \Psi(x,u,t) $ is multiplicatively renormalized if its coefficient
function satisfies the following equation
\begin{equation}
\lambda\psi(z_{1},\ldots,z_{n})=
\sum_{i<k}{\tilde H}_{ik}
\psi(z_{1},\ldots,z_n)=
1/2\sum_{i<k}({\tilde h}_{ik}
+{\tilde {\bar h}}_{ik})
\psi(z_{1},\ldots,z_n),
\label{eqcon}
\end{equation}
where ${\tilde h}_{ik}$ again has a two--particle form
\begin{equation}
{\tilde h}_{ik}\psi(..z_{i}..z_{k}..)=
\int^1_0
\prod_{m=1}^{3}{\rm d}\alpha_m
\delta (\sum_{n=1}^{3}\alpha_n - 1)
\alpha_2^{\mu - 3}\psi(..,\alpha_1 z_{i} + {\bar \alpha_3}z_{k},..,
{\bar \alpha_1}z_{i} + \alpha_3 z_{k},..) \ .
\label{psi}
\end{equation}
Again on the rhs of this equation the $i$--th and
$k$--th arguments are replaced in the manner shown. The expression
for ${\tilde {\bar h}}_{ik}$ is obtained from (\ref{psi}) by
interchanging
$ z_{i}\leftrightarrow z_{k} $.
The critical exponent of the corresponding eigenoperator is
\begin{equation}
x=n(\mu-1)+l+n\frac{\eta}{2}+\frac{2\nu(\mu)}{N}\lambda
+O(1/N^2)
\label{critdim}
\end{equation}
corresponding to a full scaling dimension
$ y=d-x$.
Here and in the sequel $ n $ is the number of fields
$ \phi$ and
$ l $ is the number of derivatives in the eigenoperator.
The function $\nu(\mu)$ has been defined in~(\ref{def_nu})
and $\eta$ was given in~(\ref{adim_phi}).

Eq.~(\ref{eqcon}) has a number of nice properties.
But the eigenvalue problem is more manageable if
written in terms of the
$ \hpsi(z_{1}..z_{n}) $ functions.
Under the
transformation~(\ref{hatpsi}) the lhs and rhs of~(\ref{eqcon})
transform into
\begin{equation}
\lambda\hpsi(z_{1},\ldots,z_{n})=\sum_{i<k}{H}_{ik}
\hpsi(z_{1},\ldots,z_n)\equiv H\hpsi(z_{1},\ldots,z_n)
\label{eqhpsi}
\end{equation}
with the operator
$ H_{ik} $ defined in Eq.~(\ref{R1}).
Some technical details of this calculation are given in Appendix~A.

Now it is easy to show that the symmetry group of~$ H $ is
$ SL(2,C) $.
We define the action of the
$ SL(2,C) $ group on functions
$ \hpsi(z_{1},\ldots, z_{n}) $ in the following manner
\begin{equation}
\rm{S}(g)\hpsi(z_1,\ldots,z_n) =
\prod_{i=1}^n (c z_i + d)^{-\Delta}\:
\hpsi\left(\frac{a z_1 + b}{c z_1 + d},\ldots,
\frac{a z_n + b}{c z_n + d}\right)
\end{equation}
where $g\in SL(2,C),\> g=\left (\begin{array}{cc} a & b \\
 c & d \end{array} \right )$ and
$a d - b c = 1$.
It is easy to check that  $H$ and
$\rm{S}(g)$ commute for all $g$
with $\Delta = \mu - 1$ in our case.
The group $\rm{SL}(2)$ has three generators
$\rm{S}\ ,\ \rm{S_{+}} \ ,\ \rm{S_{-}}$:
\begin{equation}
\rm{S}=
\sum_{i=1}^{n}
(z_i \df_{z_i} + \Delta/2),\>\>
\rm{S_{-}}=
-\sum_{i=1}^{n}\df_{z_i},\>\>
\rm{S_{+}}=
\sum_{i=1}^n(z_i^2 \df_{z_i} + \Delta z_i).
\label{generators}
\end{equation}
All these generators commute with $H$ and obey the
commutation relations
\begin{equation}
[\rm{S},\rm{S_{-}}] = - \rm{S_{-}}, \ \
[\rm{S},\rm{S_{+}}] = + \rm{S_{+}}, \ \
[\rm{S_{+}},\rm{S_{-}}] = 2\rm{S}.
\label{comm}
\end{equation}
Using Eqs.~(\ref{comm}) it is straightforward to see that the
eigenvalue problem for $ H $ can be restricted
to the subspace of functions annihilated by ${\rm S_-}$:
\beq
S_{ln}^{-}=\{\hpsi(z_{1},\ldots,z_{n})|
\mbox{${\rm S}_{-}\hpsi=0;\ $}
\mbox{$\rm{S}\hpsi=(l+n\Delta/2)\hpsi\}$} \ .
\eeq
According to (\ref{transl}) the operators in ${\rm S_-}$
are just the conformal operators transforming like~(\ref{def_conf}).
Like in Ref.~\cite{KWP} we therefore observe that the spectrum of critical
exponents is generated by the conformal operators.
All other eigenfunctions
of $ H $ are obtained by application of $ \rm{S}_{+} $ on the
eigenfunctions in $ S_{ln}^{-}, \>(l,n=0,1,\ldots)$.  Thus we have
to solve the eigenvalues problem~(\ref{eqhpsi}) for the
operator $ H $  on the space of functions $\hpsi(z_1,...,z_n)$
that are symmetric, translation invariant and
homogeneous polynomials of degree $l$ in $n$ variables.

Another very useful property of
$ H $ is its hermiticity with respect to a suitable scalar product
defined on the space of homogeneous polynomials as
\begin{equation}
\label{scalar}
<\hpsi_{1}|\hpsi_{2}> \equiv
\psi_{1}(\df_{z_1},
\ldots,\df_{z_n})
\hpsi_{2}(z_1,\ldots,z_n)
\bigl|_{z_i = 0} \ .
\end{equation}
Remember that
$ \psi $ and
$ \hpsi $ are related via transformation~(\ref{hatpsi}).
To prove the hermicity of
$ H $ with respect to the scalar product~(\ref{scalar})
\begin{equation}
\label{hermicity}
<\hpsi_{1}|H\hpsi_{2}> =<H\hpsi_{1}|\hpsi_{2}> \ ,
\end{equation}
it is sufficient to note that
$ {\tilde H}_{ik}\psi(z_{1},\ldots,z_n)\to
{H}_{ik}\hpsi(z_{1},\ldots,z_n)$ under the transformation~(\ref{hatpsi}).

An important immediate consequence of this hermicity
is the reality of the spectrum of critical exponents, i.e. all
anomalous dimensions of eigenoperators in~${\cal C}_{\rm sym}$
are real numbers at least in first order in~$1/N$.

\subsection{Double expansions}
A closed analytical solution of Eq.~(\ref{eqhpsi}) seems not possible
except for the trivial case
$ n=2 $ where the form of the eigenfunctions is already
fixed by conditions imposed on the functions
$ \hpsi(z_{1},z_{2}) $ alone. Indeed there is only one symmetric
translation invariant polynomial of degree
$ l=2L $:
$ \hpsi_{l}(z_{1},z_{2})=(z_{1}-z_{2})^{l} $. The corresponding
eigenvalue of (\ref{eqhpsi}) is
$ \lambda_{l}=1/((l+\mu-2)(l+\mu-1)) $ leading to critical
exponents~(\ref{critdim}) in agreement with known results~\cite{LR2}.

There are two important cases when the situation is simplified
and the problem becomes more tractable. Let us consider the eigenvalue
problem in the limiting cases where
$ d=2+\eps $ or $ d=4-\eps $, in other words we want to hold the
leading terms in
$ \eps $ in the anomalous dimensions.
This corresponds to a double expansion in~$1/N$ and~$\eps$ keeping
the leading nontrivial terms in Eq.~(\ref{critdim}).
The multiplier $ \nu(\mu) $ is in both cases
proportional to $ \eps^{2} $, whereas
$ \lambda $ determined from Eq.~(\ref{eqhpsi}) has a simple pole in
$ \eps $.  Extracting the pole contribution from the expression
$ H\hpsi(z_{1},\ldots,z_{n}) $
(defined by Eq.~(\ref{R1})) one obtains the following
formulae for the critical exponents with
$ O(\eps^{2}) $ accuracy in
$ d=2+\eps $
\begin{equation}
x=n\frac{\eps}{2}\left(1+\frac{1}{N}\right)+l+
\frac{\eps}{2N}\lambda_{2+\eps}
+O(\eps^2)+O(1/N^2)
\label{twoplusx}
\eeq
and $d=4-\eps$ dimensions
\beq
x=n\left(1-\frac{\eps}{2}\right)+l+\frac{\eps}{N}\lambda_{4-\eps}
+O(\eps^2)+O(1/N^2) \ ,
\label{fourminusx}
\end{equation}
where $\lambda_{2+\eps}$ and $\lambda_{4-\eps}$  are given by
the eigenvalue problems
\begin{equation}
\lambda_{2+\eps}\hpsi(z_{1},\ldots, z_{n})=\sum_{i<k}
\left(
\hpsi(..,z_{i},\ldots, z_{i},..)+\hpsi(..,z_{k},\ldots, z_{k},..)
\right)
\label{twoplus}
\end{equation}
and
\begin{equation}
\lambda_{4-\eps}\hpsi(z_{1},\ldots, z_{n})=2\sum_{i<k}
\int_{0}^{1}d\alpha\hpsi(..,z_{i}\alpha+(1-\alpha)z_{k},\ldots,
z_{i}\alpha+(1-\alpha)z_{k},..) \ .
\label{fourminus}
\end{equation}
Again the $i$--th and
$k$--th arguments on the rhs of these equations are replaced
in the specified manner.

These double expansions can be compared with known results,
which serves as a consistency check of our calculations.
Formula~(\ref{fourminus}) has already been obtained in
Ref.~\cite{KW} in a $4-\eps$~expansion. In fact, one can
also check in a lengthy calculation that the term in
order~$\eps^2/N$ from Ref.~\cite{K} is in agreement
with our $1/N$--results.

In the $2+\eps$~expansion the complete spectrum of critical
exponents in order~$\eps$ has been obtained by Wegner~\cite{W1}.
The solution of the eigenvalue problem~(\ref{twoplus}) is
straightforward and yields
\beq
\lambda_{2+\eps} = m(2n-m-1),\qquad m=0\ldots n \ .
\label{twoplusspectrum}
\eeq
Here $m$ is the maximum number of elementary fields without any
gradients acting on them in any term of
the respective eigenoperator.
This is in agreement with the results of Wegner~\cite{W1}
restricted to our subalgebra~${\cal C}_{\rm sym}$ as can be
checked easily.

\section{An asymptotic naive addition law for anomalous dimensions}
\setcounter{equation}{0}
\subsection{Outline of the proof}

Since we have the exact solution of the eigenvalue problem for
operators from ${\cal C}_{\rm sym}$ in first order in
the $ 2+\eps $ expansion,
it is natural to try to understand how large the
corrections from higher orders in~$\eps$ are.
Indeed, the calculation of critical exponents of high--gradients
operators in ${\cal C}_{\rm sym}$
will lead to the observation that an appropriate expansion
parameter is not $ \eps $
but some combination of $ \eps $ and~$ l $.
Later we will see that a suitable combination is
e.g. $ \eps\cdot\ln{l} $.

The $ 1/N$--expansion, in which the higher orders in~$\eps $
are summed up, is very suitable for this problem.
However, due to very fast growth of the size of the mixing matrix
with increasing $ l $
it seems hardly possible to obtain the analytical solution of
the eigenvalue problem Eq.~(\ref{eqhpsi}).
But the following interesting quantitative prediction for critical
exponents of high--gradient operators can be done.

\vspace*{0.3cm} \noindent
{\bf Theorem:~~}
Let
$ \Lambda^{l}_{n} $ be the spectrum of
$ H=\sum_{i<k}H_{ik} $ restricted to the subspace
$ S_{-}^{n,l} $, i.e. the subspace of symmetric homogeneous polynomials
of degree $ l $ in $ n $ variables. Then for any two eigenvalues
$\lambda_{1}\in \Lambda^{R_{1}}_{n_{1}}$
and
$\lambda_{2}\in \Lambda^{R_{2}}_{n_{2}}$
corresponding to different conformal eigenoperators
the following statement holds:
There is a number~$L$ such that for all\footnote{
If $\lambda_1$ and $\lambda_2$ correspond to the same eigenoperator,
this statement holds only for even~$l>L$.} $l>L$
one can find an eigenvalue
$\lambda_{l}\in \Lambda^{l}_{n_1+n_2}$\
such that
\begin{equation}
|\lambda_{l}-\lambda_{1}-\lambda_{2}|\leq C/l^{{\Delta}} \ .
\label{uneq}
\end{equation}
Here $ C $ is a fixed constant
and $ \Delta=\mu-1 $.

\vspace*{0.3cm}
Hence any two anomalous dimensions for composite operators with
$n_1$ and $n_2$ elementary fields generate a limit point
given by the sum of these anomalous dimensions
in the spectrum of composite operators with
$n=n_1+n_2$ fields. The approach to this limit point is
given by~(\ref{uneq}). Therefore Eq.~(\ref{uneq}) is an
``asymptotic naive addition law" for anomalous dimensions.
The limit point structure in the spectrum leads to a clustering
of anomalous dimensions. For more details we refer the reader
to Refs.~\cite{K,DM1} where this property was discussed
in the framework of the $4-\eps$~expansion.

The general scheme of the proof of the theorem is the same as in the
paper~\cite{DM1}.
So here we will only discuss the main ideas
and some technical modifications as compared to~\cite{DM1}.
The full proof can then be constructed along the same lines as
in Ref.~\cite{DM1}.

To prove the unequality~(\ref{uneq}) it is sufficient
to show that in the subspace
$ S_{-}^{l,n} $  there exists a vector
$ \hPsi_{S}(z_{1},\cdots,z_{n})  $ such that
\begin{equation}
||(H-\lambda_{1}-\lambda_{2})\hPsi_{S}||/||\hPsi_{S}||\leq
C/l^{{\Delta}} \ .
\label{uneqnorm}
\end{equation}
Here, as usual,
$ ||\hPsi_{S}||^{2}=<\hPsi_{S}|\hPsi_{S}> $.

We begin with the construction of the test vector
$ \hPsi_{S} $. If
$ \hpsi_{1}(z_{1},\cdots,z_{n_{1}}) $
and
$ \hpsi_{2}(z_{1},\cdots,z_{n_{2}}) $ are the eigenfunctions of
$ H $ corresponding to the eigenvalues
$ \lambda_{1} $ and
$ \lambda_{2} $, we define
$ \hPsi_{S} $ in the following way
\begin{equation}
\hPsi_{S}(z_{1},\cdots,z_{n})=
\mbox{\bf T}\Psi_{S}(z_{1},\cdots,z_{n})=
{\mathop{\mbox{\bf Sym}}_{\scriptstyle{\{z_{1},\ldots,z_{n}\}}}}
\mbox{\bf T}\Psi(z_{1},\cdots,z_{n})\ ,
\label{symform}
 \end{equation}
where $\mbox{\bf Sym}$
denotes  symmetrization with respect to  $z_{1},\cdots,z_{n}$.
The function $ \Psi(z_{1},\cdots,z_{n})$ reads
\begin{equation}
\Psi(x,y)=\sum_{k=0}^{l}c_{k}
(x_{1}+\cdots+x_{n_{1}})^{k}
(y_{1}+\cdots+y_{n_{2}})^{l-k}\psi_{1}(x)\psi_{2}(y)\ ,
\label{fundef}
\end{equation}
where $ c_k=(-1)^{l-k}\bin{l}{k}\Gamma(S_{1})\Gamma(S_{2})/
\Gamma(k+S_{1})\Gamma(l-k+S_{2})  $,
\mbox{$S_{i}=n_{i}\Delta+2R_{i}$} and
$\bin{l}{k}$ is the binomial coefficient.
Also, for brevity, we use the shorthand notation
$\psi_{1,2}(x)=\psi_{1,2}(x_{1},\cdots,x_{n_{1,2}})$ and
$\Psi(x,y)=\Psi(x_{1},\cdots,x_{n_{1}},y_{1},\cdots,y_{n_{2}})$.

A straightforward calculation shows that
$ \Psi(z) $  defined by Eq.~(\ref{fundef}) satisfies
Eq.~(\ref{confeq}). Consequently, the functions
$ \hPsi(z) $, $ \hPsi_{S}(z) $ are translation invariant.
The degree of the polynomial $ \hPsi_{S}(z) $ is obviously equal to
$ l+R_{1}+R_{2} $.

To prove the unequality~(\ref{uneqnorm}) for
$ \hPsi_{S} $ given by Eq.~(\ref{symform}) one needs to calculate
$ ||\hPsi_{S}|| $ and
$ ||(H-\lambda_{1}-\lambda_{2})\hPsi_{S}||$.
To do this we write the functions
$ \Psi_{S} $ and
$ \hPsi_{S} $ in the following more suitable form
\begin{equation}
\Psi(x,y) =
\psi_{1}(x)
\psi_{2}(y)
K(\df_{a},\df_{b})
\exp\{a(\sum_{i=1}^{n_1}x_i)
+ b(\sum_{j=1}^{n_2}y_j)\}
\biggl|_{a,b = 0}\ ,
\label{block1}
\end{equation}
where
$ K(a,b)=\sum_{k=0}^{l}c_{k}a^{k}b^{l-k} $.

Respectively, applying the transformation~(\ref{hatpsi})
to Eq.~(\ref{fundef}) yields after some algebra
\begin{equation}
\hPsi(x,y) =
\int_{0}^{1}d\al d\bet \int_{0}^{1} Du Dv
f(\al,n_{1},R_{1})
f(\bet,n_{2},R_{2}) [{\bar \al}(ux)-{\bar\bet}(vy)]^{l}
\psi_{1}(ux)\psi_{2}(vy)\ ,
\label{block2}
\end{equation}
where
\mbox{$ f(\al,n,R)=(1-\al)^{S-R-1}\al^{R-1} \cdot \Gamma(S)/\Gamma(R)
\Gamma^{n}(\Delta)$};
$ Du=\prod_{i=1}^{n_{1}}du_{i}u_{i}^{\Delta-1}\delta(\sum_{i}u_{i}-1)$;
$ ux\equiv\sum_{i}u_{i}x_{i} $
and $ \psi(ux)=\psi(u_{1}x_{1},\ldots,u_{n_{1}}x_{n_{1}}) $.

Now the calculation of the norm of
$ \hpsi_{S} $ can be done in the following manner:\\
1. Using formulae~(\ref{scalar}), (\ref{symform}) and taking into
account that the functions
$ \psi_{1,2}(z) $ are fully symmetric one obtains
\begin{equation}
||\psi^{l}_{S}||^{2}=
(\bin{n_{1}+n_{2}}{n_{1}})^{-1}
\sum_{k=0}^{n_{2}}\bin{n_{1}}{k}\bin{n_{2}}{k}A^{(k)}_{l}\ ,
\label{scprod}
\end{equation}
where
$A^{(k)}_{l}=\Psi(\df_x,\df_y)\hPsi(y_{1},\ldots,y_{k},x_{k+1},
\ldots,x_{n_{1}},x_{1},\ldots,x_{k},y_{k+1},\ldots,y_{n_{2}})$.
Without loss of generality we now assume
$ n_{2}\leq n_{1} $.\\
2. Using formulae~(\ref{block1}), (\ref{block2}) and the translation
invariance of
$ \hPsi $ one can obtain the following expression for
$ A^{(k)}_{l} $
\begin{equation}
A^{(k)}_{l}=N(l)
\int_{0}^{1}d\al d\bet \int_{0}^{1} Du Dv
f(\al,n_{1},R_{1})
f(\bet,n_{2},R_{2}) {\cal A}^{l}(\al,\bet,u,v){\cal F}(\al,\bet,u,v)\ ,
\label{blockA}
\end{equation}
where
${\cal A}(\al,\bet,u,v)=
[{\bar\al}
({\bar U}_{k}-{U}_{k})+
{\bar\bet}
({\bar V}_{k}-{V}_{k})]/2
$;
$ {U_{k}}=\sum_{i=0}^{k}u_{i}$;
$ {V_{k}}=\sum_{i=0}^{k}v_{i}$;\\
$ N(l)=K(\df_{a},\df_{b})(a-b)^{l}|_{a=b=0} $ and
$$
{\cal F}(\al,..)=\sum_{i=0}^{R_{1}+R_{2}}\bin{l}{i}
{\cal A}^{-i}
\df_{a}^{i}
\psi_{1}(\df_x-a{\bar \bet}v,
\df_x-a{\bar \al}u)
\psi_{2}(\df_y+a{\bar \al}u,
\df_y-a{\bar \bet}v)
\psi_{1}(uy,
ux)
\psi_{2}(vx,
vy)\biggl|_{\stackrel{{\scriptstyle{x=1/2}}}{y=-1/2}}.
$$
Here
$ \psi_{i}(x,y) $ means
$ \psi_{i}(x_{1},..,x_{k},y_{k+1},..,y_{n_{i}})$ and we mention
again that $ {\bar x}\equiv(1-x) $.\\
3. All the dependence on
$ l $
in expression~(\ref{blockA}) for
$ A^{(k)}_{l} $, except the trivial one contained in the multiplier
$ N(l) $ (we do not need its explicit form), is due to the factor
${\cal A}^{l}$
in the integral. It is evident that the main contributions
in the integral come from the regions in which
$ |{\cal A}(...)|\simeq 1 $:
$$ \{ {\cal A}\simeq 1
\Longleftrightarrow
\al \simeq \bet \simeq
U_{k} \simeq
V_{k} \simeq 0;
\}; \>\>\>
\{{\cal A}\simeq -1
\Longleftrightarrow
\al \simeq \bet \simeq
{\bar U}_{k} \simeq
{\bar V}_{k}
\simeq 0\}\ .
$$
Then after long but straightforward calculations one finds that
the leading contributions in the integral~(\ref{blockA})
from these regions behave as
$ C_{1}^{(k)}/l^{2k\Delta} $ for
$ A\simeq 1 $ and
$ C_{-1}^{(k)}/l^{(n_{1}+n_{2}-2k)\Delta} $ for
$ A\simeq -1 $. The explicit expressions for the
constants $C_{\pm 1}^{(k)}$ are
\begin{equation}
C_{1}^{(k)}=
\frac{\Gamma(S_{1})\Gamma(S_{1})}{\Gamma^{n_{1}+n_{2}-2k}(\Delta)}
\int [D^{n_{1}}_{k+1}u]\ \psi_{1}^{2}(0,u) \:
\int [D_{k+1}^{n_{2}}v] \ \psi_{2}^{2}(0,v),
\end{equation}
\begin{equation}
C_{-1}^{(k)}= (-1)^{l}
\frac{\Gamma(S_{1})\Gamma(S_{1})}{\Gamma^{2k}(\Delta)}
\int [D^{k}_{1}u]\
(\psi_{1}(u,0)\psi_{2}(u,0)) \:
\int [D^{k}_{1}v]\ (\psi_{1}(v,0)\psi_{2}(v,0)),
\end{equation}
where $[D_{k}^{n}u]\equiv
\prod_{i=k}^{n}du_{i}\,u_{i}^{\Delta-1}\,\delta(\sum_{k}^{n} u_{i}-1)$.

Thus one can see that the main contribution to the norm of
$ \hPsi_{S} $ is given by $A_{l}^{(k)}$
when
$ k=0 $  or
$ 2k=n_{1}+n_{2} $. It can be shown that in these cases the
\underline{exact} answer for $A_{l}^{(k)}$  can be obtained and it
coincides with the answer for the leading term of the asymptotic
behaviour of $A_{l}^{(k)}$. Put in another way, the functions
$ \hPsi(z) $ with arguments arranged in different ways are
almost orthogonal to one another for large $ l $.

Eventually the expression for the norm of
$ \hPsi_{S} $ is
\begin{equation}
||\hPsi_{S}||^{2}=N(l)(\bin{n_{1}+n_{2}}{n_{1}})^{-1}
||\hpsi_{1}||^{2}||\hpsi_{2}||^{2}
(1+(-1)^{l}\delta_{\psi_{1},\psi_{2}})(1+O(1/l^{\Delta}))\ .
\label{normpsi}
\end{equation}
\vskip 0.5cm

Our next task is to obtain an estimate from above for
${\cal E}(l)\equiv ||(H-\lambda_{1}-\lambda_{2})\hPsi_{S}||$.
Instead of trying to derive the exact asymptotic behaviour for
this quantity, we prefer to give a rough estimate.
This, however, reproduces the true dependence on
$ l $ but in a much shorter calculation.\\
1.\  First of all, taking into account that
$ H=\sum_{i<k}H_{ik} $ is symmetric under transpositions of
its arguments
and using Schwartz's inequality one gets
\begin{equation}
{\cal E}(l)=||(H-\lambda_{1}-\lambda_{2})\hPsi_{S}||\leq
||(H-\lambda_{1}-\lambda_{2})\hPsi|| \ .
\end{equation}
Remember that
$ \hPsi(z)=\hPsi(z)(z_{1}\cdots z_{n} $.\\
2.\ It can be easily shown that
$ (z_{1}+...z_{n})\psi(z)\to {\rm S}_{+}\hpsi(z)$ under
transformation~(\ref{hatpsi}). Then taking advantage of the explicit
form of
$ \Psi(z) $~(Eq.~(\ref{fundef})) and the
$ SL(2,C) $ invariance of
$ H $ one obtains:\\
$
\sum_{i<k}^{n_{1}}H_{ik}\hPsi(z)=\lambda_{1}\hPsi(z) \>
\mbox{ and } \>
\sum_{n_{2}<i<k}H_{ik}\hPsi(z)=\lambda_{2}\hPsi(z).
$
Taking into account the symmetry properties of
$ \hPsi(z) $ and again using Schwartz's inequality we find
\begin{equation}
 {\cal E}(l) \leq n_{1}n_{2}<\hPsi(x,y)H^{2}(x_{1},y_{1})\hPsi(x,y)>\ .
\label{estimation}
\end{equation}
It is possible to simplify Eq.~(\ref{estimation}) even more.
Let us remember that the eigenvalues of
$ H_{ik}=1/2(h_{ik}+{\bar h}_{ik}) $ are
$ \lambda_{l}=[(l+\mu-2)(l+\mu-1)]^{-1} $
$ l=0,2,4.. $. One sees that for
$ 1<\mu<2 $ the only negative
eigenvalue is $ \lambda_{0} $. All the other eigenvalues lie in the
interval
$ (0,1) $. Thus if
$ P_{0} $ is the projector onto the subspace of eigenfunctions of
$ H(x,y) $ with
$ \lambda=\lambda_{0} $ one has
$ <\hPsi|H(x_{1},y_{1})|\hPsi>\leq$
\mbox{$ <\hPsi|(H-\lambda_{0}P_{0})|\hPsi>$}
\mbox{$+ \lambda_{0}^{2}<\hPsi|P_{0}|\hPsi> \leq$}
\mbox{$<\hPsi|{\bar h}(x_{1},y_{1})|\hPsi>+\lambda_{0}(\lambda_{0}-1)
<\hPsi|P_{0}|\hPsi>$}.\\
The projector
$ P_{0} $ can be written as
\begin{equation}
 P_{0}\hpsi(x,y)=B^{-1}(\Delta,\Delta)\int_{0}^{1}ds(s(1-s))^{\Delta-1}
 \hpsi(sx+(1-s)y,sx+(1-s)y)\ .
 \end{equation}
Here
$ B(x,y) $ is Euler's beta function.\\
3.\ The calculation of
$ <\hPsi|P_{0}|\hPsi>  $
and
$ <\hPsi|{\bar h}|\hPsi>  $
is carried out in a similar manner, so we consider the latter
only. First of all it is useful to rewrite the expression~(\ref{R1})
for
$ {\bar h}_{ik} $
in a well--defined form for
$  2<d<4 $~(compare Eq.~(\ref{R1}) and the comments to it).
\begin{equation}
{\bar h}_{ik}\hpsi(x,y)=
(\Delta-1)^{-1}\int_{0}^{1}ds\,dt\,
t(1-t)^{\Delta-1}\,
(\mu+t\df_t)\,
\hpsi((1-st)x+sty,t{\bar s}x+(1-{\bar s}t)y).
\end{equation}
Furthermore, in full analogy with the derivation of Eq.~(\ref{blockA})
we obtain
\begin{eqnarray}
\lefteqn{<\hPsi|{\bar h}(x_{1},y_{1})|\hPsi>=\frac{N(l)}{\Delta-1}
\int_{0}^{1}d\al\, d\bet\, ds\, dt\,  Du\, Dv\:
f(\al,n_{1},R_{1})
f(\bet,n_{2},R_{2})}\ret
&&\times\: t(1-t)^{\Delta-1}\,
(\mu+t\df_t)\,
{\cal A}^{l}(\al,\bet,u,v,t,s){\cal G}(\al,\bet,u,v,t,s)\ ,
\label{blockB}
\end{eqnarray}
where
${\cal A}=[{\bar \al}(1/2-u_{1}ts)+{\bar\bet}(1/2-v_{1}t{\bar s})]$.
The function
${\cal G} $
is a polynomial of degree
$ R_{1}+R_{2} $
and has a structure like
$  {\cal F} $ in Eq.~(\ref{blockA})
Again one concludes that the dominant contribution to the integral
comes from the region where
$ |{\cal A}|\simeq 1$. At first sight the terms arising from
$ t\df_t $ in~(\ref{blockB}) give the leading term of the asymptotic
behaviour as
$ l\to \infty $, since differentiation
of
$ {\cal A}^{l} $ produces additional power of
$ l $. But these additional powers of
$ l $ come in pair with ``small variables"
$ \sim (1-{\cal A}) $ that lead to the observation that the
``$ \mu $--term" and the ``$t\df_t$--term" in Eq.~(\ref{blockB})
have the
same order in $ l $. Both these terms can be treated on an
equal footing.
We consider the first of them. The condition
$ 1-{\cal A}\leq \vareps $  leads to
$ \al\simeq\bet\simeq u_{1}ts\simeq v_1 t{\bar s}\leq \vareps $. Then
in this region
$ {\cal A}\simeq (1-\al/2)(1-\bet/2)(1-u_{1}ts-v_{1}t{\bar s}))
+O(\vareps^{2}) $.
Function
$ {\cal G} $ in its turn
can be bounded by
$ Const\cdot l^{R_{1}+R_{2}} $. Using this the  integration
with respect to
$ u_{2},..u_{n_{1}},v_{2},..v_{n_{2}} $ and
$ \al,\bet $ can be performed easily
giving a factor $ l^{-(R_{1}+R_{2})} $.
Thus the integral under consideration can be bounded from above
by the following expression
$$
C(\Delta)N(l)\int_{0}^{1}du\,dv\,ds\,dt\:
t(1-t)^{\Delta-1}
{\bar u}^{(n_{1}-1)\Delta-1}
{\bar v}^{(n_{2}-1)\Delta-1}
(1-ust-vt{\bar s}))^{l}\ .
$$
After tedious but straightforward calculations one shows that the leading
term of the asymptotic expansion of this integral behaves as
$ l^{-2\Delta} $. So the final answer for
$ {\cal E}(l) $ is
\begin{equation}
{\cal E}(l) \leq C N(l)l^{-2\Delta}.
\label{normH}
\end{equation}
It is clear that the unequalities~(\ref{uneqnorm}),~(\ref{uneq})
now follow immediately from
Eqs.~(\ref{normpsi}), (\ref{normH}).

\subsection{Failure of the limit point structure in the
$2+\eps$~expansion}
{}From the results in the previous subsection it is clear
that the approach to the limit point close to two dimensions
is like $l^{-\eps/2}$, where $l$ is the total number of gradients.
Hence the behaviour is non--uniform for $d\in [2,4]$ and only
uniform for $d\in [2+\delta,4]$ with some fixed~$\delta>0$.
A suitable expansion parameter for high--gradient operators
close to two dimensions is therefore necessarily some
combination of~$\eps$ and~$l$, e.g. the most straightforward
choice is~$\eps\cdot\ln l$.

Let us make this more explicit by looking at the spectrum
of conformal operators with $n=3$~fields. The anomalous dimensions
obtained by diagonalizing~(\ref{eqhpsi}) are plotted in Fig.~3.
Let us concentrate on the curves approaching the limit
curve labeled~$\phi^2\otimes\phi$ given by the (trivial) sum
of anomalous dimensions of~$\phi^A$ and
$h_{A_1 A_2}\phi^{A_1}\phi^{A_2}$. Here $h_{A_1 A_2}$ is a symmetric
and traceless tensor. First of all one can readily check that
the approach to the limit curve as a function of~$\l$ is
in fact like~$l^{-d/2+1}$, with even~$l$ approaching from
above and odd~$l$ from below. The approach is clearly non--uniform
in the interval~$d\in [2,4]$. Close to two dimensions the
derivatives of the curves even diverge for
$l\rightarrow\infty$, corresponding to diverging terms
in order~$\eps^2/N$ in the $2+\eps$~expansion.

That the limit point structure cannot be observed in the
$2+\eps$~expansion is also apparent from the diagonalization
of~(\ref{twoplus}). We find the following possible eigenvalues for
different numbers of fields~$n$ according to~(\ref{twoplusspectrum}):
\beqar
n=1\ &:&\qquad \lambda_{2+\eps}=0 \ret
n=2\ &:&\qquad \lambda_{2+\eps}=0,2 \ret
n=3\ &:&\qquad \lambda_{2+\eps}=0,4,6 \nonumber
\eeqar
This of course agrees with the values in Fig.~3 for $d=2$.
The naive sum of the eigenvalues belonging to $\phi^A$ and
$h_{A_1 A_2}\phi^{A_1}\phi^{A_2}$ with $n=1,2$ is
$\lambda_{2+\eps}=2$, but this value is not contained in the
above list of possible eigenvalues with $n=3$~fields.
Similar observations can be made quite generally in the spectrum
of the $2+\eps$~expansion, i.e. the limit point structure
(\ref{uneq}) remains invisible.

\setcounter{equation}{0}
\section{Vasil'ev and Stepanenko's results re--examined}

\subsection{Interpolation of the high--gradient behaviour for $2<d<4$}
In the operator subalgebra~${\cal C}_{\rm sym}$ discussed so far
we have seen that one must be cautious if one uses~$\eps$ as
an expansion parameter in~$d=2+\eps$ for high--gradient operators.
This can lead to wrong conclusions about the limit of a
large number of gradients~$l\rightarrow\infty$. Now, however,
the problematic class of high--gradient operators giving rise
to Eqs.~(\ref{hg1}),~(\ref{hg2}) does not belong to~${\cal C}_{\rm sym}$.
The structure of the respective eigenoperators in the
$2+\eps$~expansion is~\cite{W1}
\beq \label{hg3}
\sum_{k=0}^l \gamma_k
(\partial_+\vec\pi\cdot\partial_-\vec\pi)^{2(l-k)}
(\partial_+\vec\pi\cdot\partial_+\vec\pi)^k
(\partial_-\vec\pi\cdot\partial_-\vec\pi)^k
\eeq
with some coefficients~$\gamma_k$ ($\gamma_0\neq 0$) and
$\partial_\pm=\frac{1}{\sqrt{2}}(\partial_x\mp i\partial_y)$.
Hence these eigenoperators of the nonlinear $\sigma$--model
(\ref{sigmamodel}) are $O(N)$ and $O(d=2)$--scalar.

In this section we use the opportunity to show that the
operators~(\ref{hg3}) have in fact already been investigated
in the first order of the $1/N$--expansion for $2<d<4$
by Vasil'ev and Stepanenko~\cite{VS}. As this is not immediately
apparent from their work, we want to clarify this point here.

In Ref.~\cite{VS} Vasil'ev and Stepanenko calculated the
critical exponents of the composite operator~$\sigma^s$
made up of powers of the auxiliary field~$\sigma$
from~(\ref{action}).\footnote{In Ref.~\cite{VS} the auxiliary
field was denoted by~$\psi$.} They find the following
full scaling dimension
\beq \label{hg1N}
y=d-2s-\eta\,s(d-1)
\frac{(s-1)d(d-3)-2(d-1)}{4-d}+O(N^{-2})
\eeq
with $\eta$ from (\ref{adim_phi}).
If one expands this additionally in $d=2+\eps$ one finds
\beq \label{1N2eps}
y=d-2s+\frac{\eps}{N}s(s-1)+O(\eps^2)+O(N^{-2})
\eeq
or alternatively in $d=4-\eps$
\beq \label{1N4eps}
y=d-2s-6\frac{\eps}{N}s(s-2)+O(\eps^2)+O(N^{-2}) .
\eeq
One readily observes that (\ref{1N2eps}) corresponds to the
high--gradient operators~(\ref{hg3}) as the critical exponent
is consistent with~(\ref{hg1}). In $d=4-\eps$ Eq.~(\ref{1N4eps})
corresponds to the composite operators~$(\vec\phi^2)^s$ with
the full scaling dimension~\cite{W3}
\beq
y=d-2s-\eps\,\frac{6s(s-2)}{N+8} +O(\eps^2) .
\eeq
As these critical exponents are unique in the respective
expansions, this proves that powers of the auxiliary field~$\sigma^s$
interpolate between the high--gradient operators~(\ref{hg3})
with $2s$~fields and $2s$~gradients in $d=2+\eps$ and
the composite operators~$(\vec\phi^2)^s$ with $2s$~fields and
no gradients in~$d=4-\eps$. This might appear surprising
at first sight as both~$\sigma^s$ and~$(\vec\phi^2)^s$ contain
no gradients. However, the canonical dimension of~$\vec\phi$ is
$x_\phi=d/2-1$, which varies as
\beqar
d=4-\eps & : & x_\phi=1-\eps/2 \ret
d=2+\eps & : & x_\phi=\eps/2 \nonumber
\eeqar
and hence allows to built gradients into the respective
eigenoperators as~$d$ varies from four to two. Therefore the
``most stable" eigenoperators~$(\vec\phi^2)^s$ in $d=4-\eps$
are linked with the unstable high--gradient operators~(\ref{hg3}).

Eq.~(\ref{hg1N}) allows to answer the question where this
behaviour bends over between two and four dimensions.
Eq.~(\ref{hg1N}) can be written in the form
\beq \label{hg1NK}
y=d-2s-\frac{1}{N}\Big(s(s-1)\,K(d)+O(s)\Big)+O(N^{-2})
\eeq
with
\beq
K(d)=4(d-3)\, \frac{\G(d-2)}{\G^2(d/2-1)\G(2-d/2)\G(d/2)}\,(d-1)\ .
\label{Kd}
\eeq
$K(d)$ is plotted in Fig.~4. It has the
following general behaviour:
\beqar
K(d)>0 & \mbox{for} & d>3 \ret
K(d)=0 & \mbox{for} & d=3 \ret
K(d)<0 & \mbox{for} & d<3 \nonumber
\eeqar
As $K(d)$ multiplies the quadratic term in~$s$ in~(\ref{hg1NK}),
this yields an infinite number of relevant operators in first
order in~$1/N$ for $2<d<3$ in the limit~$s\gg N$.
Hence $d=3$ is the dimension where this seemingly unstable
behaviour changes into the stable behaviour of the
$4-\eps$~expansions~\cite{KWP}. Of course this observation
can be affected in any direction by the unknown terms
in order~$1/N^2$.

\subsection{Contact exponents}
As a short aside we briefly put the above observations into
perspective of a scenario suggested by Duplantier and Ludwig
\cite{DL}. They have argued that only field theories with
negative contact exponents~$\Theta\leq 0$ have the potential
to describe multifractal scaling phenomena (leaving aside the
use of analytical continuations like e.g. the replica method
in field theory). A contact exponent~\cite{C} $\Theta=x_k-x_i-x_j$
is defined from the short--distance OPE
\beq
O_i(r_1)\,O_j(r_2)=|r_1-r_2|^{x_k-x_i-x_j} O_k(r_2)
+\mbox{~subdominant terms with~}x_{k'}>x_k.
\eeq
Multifractal scaling requires \cite{DL} $\Theta\leq 0$, whereas
in field theories one generally finds $\Theta\geq 0$ \cite{DL}.
E.g. in the $4-\eps$~expansion one has for the symmetric traceless
part of $\phi^{A_1}\ldots\phi^{A_n}$ the unequality
$x_n-n\,x_1=n(n-1)\eps/(N+8)+O(\eps^2)>0$ in agreement with
$\Theta\geq 0$. Similar for $(\vec\phi^2)^s$ one finds
\beq
x_{(\vec\phi^2)^s}-s\,x_{\vec\phi^2}
=6s(s-1)\frac{\eps}{N+8}+O(\eps^2) >0 \ .
\label{contact1}
\eeq
However, as we have seen in the previous subsection the behaviour
(\ref{contact1}) changes qualitatively below three dimensions
in first order in~$1/N$. For powers of the auxiliary field
$\sigma^s$ one deduces from Ref.~\cite{VS}
\beqar
x_{\sigma^s}-s\,x_\sigma&=&\frac{1}{N}s(s-1)\,K(d)+O(N^{-2}) \ret
&=& \left\{ \begin{array}{ll}
>0 & \mbox{~for~} d>3 \\
=0 & \mbox{~for~} d=3 \\
<0 & \mbox{~for~} d<3
\end{array} \right.
\eeqar
with $K(d)$ defined in (\ref{Kd}). Hence below three dimensions
one finds the unusual behaviour for field theories
\beq
x_{\sigma^s}<s\,x_\sigma \mbox{~for~} 2<d<3
\label{contact2}
\eeq
in first order in $1/N$ allowing for negative contact
exponents~$\Theta<0$. In the framework of the $2+\eps$
expansion this was already noticed in Ref.~\cite{DL}.
Notice that in Eq.~(\ref{contact2}) one is not forced to
consider the limit $s\gg N$, (\ref{contact2}) holds for
arbitrary~$s$ and is in this sense a truly perturbative
expansion in a possibly small parameter~$\frac{s}{N}$.

Obviously, all this makes sense only if the usual nontrivial
fixed point does not become unstable for $s\gg N$,
$2<d<3$. Which cannot be answered conclusively for the time
being as we have seen in the previous subsection.

\section{Conclusions}
In this paper we have investigated properties of high--gradient
operators in the $N$--vector model. This class of operators
shows a seemingly unstable behaviour due to the large corrections
from anomalous dimensions~(\ref{hg1}), (\ref{hg2}) found in
Refs.~\cite{W1,CC1,CC2}. Therefore it was the natural programme
of our work to use an $1/N$--expansion for $2<d<4$~dimensions
to learn something about the behaviour of these operators as
a function of~$d$. Here we briefly sum up the main results.

For the subalgebra ${\cal C}_{\rm sym}$ of composite operators
introduced in Sect.~2.1 we have obtained the spectrum of
critical exponents in first order in~$1/N$ in a similar manner
as in Refs.~\cite{KWP,KW,K} for $4-\eps$~expansions: The
spectrum is encoded as a straightforward diagonalization
problem (\ref{eqhpsi}) in the space of homogeneous, symmetric
and translationally invariant polynomials. Let us mention
that the spectrum of critical exponents in the $1/N$--expansion
has been the subject of various papers by Lang and
R\"uhl \cite{LR} that aimed at finding algebraic structures
in $d$~dimensional conformal field theories. In ${\cal C}_{\rm sym}$
our formalism gives the same results as theirs based on
operators product expansions.

In Sect.~3 we proved a similar limit point structure (or:
naive asymptotic addition law for anomalous dimensions) as
seen in Refs.~\cite{K,DM} in $4-\eps$~expansions: Two
eigenoperators with anomalous dimensions $\lambda_1, \lambda_2$
and $n_1, n_2$ elementary fields generate the limit point
$\lambda_1+\lambda_2$ in the spectrum of conformal operators
with $n_1+n_2$~elementary fields. The approach to this
limit point is like $l^{1-d/2}$ where $l$ is the total number
of gradients. Such ``limit curves" for anomalous dimensions
(see Fig.~3) have already been observed in
Ref.~\cite{LR2}.\footnote{As non--perturbative structures
are always of interest in nontrivial field theories, our
$1/N$--results lead to the interesting conjecture that this limit structure
holds to all orders in the $N$--vector model.}

In the $2+\eps$~expansion, however, this limit point structure
cannot be seen as the approach is like~$l^{-\eps/2}$ if one
uses~$\eps$ as the expansion parameter (for a detailed
discussion see Sect.~3.2).
Unfortunately this observation could not solve the stability
problem of $2+\eps$~expansions as the problematic class
of operators leading to (\ref{hg2}) does
not belong to~${\cal C}_{\rm sym}$. Still, besides the
interesting non--perturbative structure also its failure
in the $2+\eps$~expansion teaches the important lesson
that $\eps$ might not be the suitable expansion parameter
for high--gradient operators. E.g. one would also be
tempted to read of a stability problem in order~$\eps^2$
in Fig.~3 close to $d=2$ for curves with even~$l$ tending
to the limit curve $\phi^2 \otimes \phi$. But this is
bent over in higher order in~$\eps$ as can be seen there
since asymptotically the limit curve is reached. Generally
one has to be careful with expansions like~(\ref{hg1}),
(\ref{hg2}) where higher orders are larger than previous
orders for $s\gg\eps/N$.

The $O(N)$ and $O(d)$--scalar high--gradient operators
leading to (\ref{hg2}) have in fact already been investigated
in an $1/N$--expansion by Vasil'ev and Stepanenko \cite{VS}.
This remained unnoticed, probably due to the fact that
powers of the auxiliary field~$\sigma$ from~(\ref{action})
without any gradients turn out to interpolate between
unstable high--gradient operators in $2+\eps$~expansions
and the unproblematic $(\vec\phi^2)^s$--operators in
$4-\eps$~expansions. This allowed the conclusion drawn from
Eq.~(\ref{hg1NK}) that a stability problem
for the nontrivial fixed point exists
in first order in~$1/N$ for $s\gg N$ below three
dimensions.

Of course one can argue that higher orders in the
$1/N$--expansion will bend this behaviour over again, at least
for some region of the $d$--$N$--plane. It is possible
that the stability problem is only an artefact of the
first order approximations and a proper treatment requires
the investigation of the full perturbation series. This
was just the situation for the limit point structure
in the $2+\eps$~expansion discussed above.

Considering the fact that other criticism of $2+\eps$
or $1/N$--expansions \cite{CH,KZ}
also remained unresolved
and the importance of $2+\eps$~expansions
in general, the present situation must be considered unsatisfactory.
One possible direction for future work could be
along the lines of Dasgupta and Lau
\cite{DL1,DL2}. Their conclusion that topological excitations
are relevant in the low temperature phase of the Heisenberg
model could be linked to the high--gradient
problem~\cite{CC2}.

Finally we pointed out in Sect.~4.2 that powers of the auxiliary
field~$\sigma$ lead to negative contact exponents~(\ref{contact2})
for $2<d<3$. Negative contact exponents are very unusual in field
theories and could be linked to multifractal scaling phenomena
as argued in Ref.~\cite{DL}. However, this interesting possibility
relies on a stable nontrivial fixed point and hence a satisfactory
solution of the above stability problem.

\vspace*{0.5cm} \noindent
This work was supported by Grant 95--01--00569a of the Russian
Fond for Fundamental Research and by INTAS Grant 93--2492--ext
and is carried out under the research program of the International
Center for Fundamental Physics in Moscow. S.K.K. acknowledges
hospitality of the Theory Group at St.~Petersburg State
University where part of this work was performed.

\vspace*{0.5cm}
\begin{flushleft} {\bf Note added:} \end{flushleft}
After completion of this work we were made aware of
Ref.~\cite{Janke} where the role of topological defects
in the $d=N=3$~Heisenberg model has been re--examined
using Monto Carlo simulations yielding different
conclusions than Refs.~\cite{DL1, DL2}. For recent
work on topological defects in two dimensional systems
see also Ref.~\cite{Zumbach}.

\vspace*{1cm}
\setcounter{section}{1}
\setcounter{equation}{0}
\noindent{\Large\bf Appendix A}
\renewcommand{\thesection}{\Alph{section}}
\renewcommand{\theequation}{\thesection.\arabic{equation}}
\label{ApA}
\vspace*{0.5cm}

Here we show that under transformation~(\ref{hatpsi})
Eq.~(\ref{eqcon}) transforms into Eq.~(\ref{eqhpsi}).
It is obvious that it
is sufficient to do this for the two particle case.
For simplicity we consider the operator ${\tilde h}_{12}$ only
\beqar
{\tilde h}_{12}\psi(z_{1},z_{2})&\!\!\!=\!\!\!&\psi_{h}(z_{1},z_{2})=
\int_{0}^{1}d\al\int_{0}^{{1-\al}}d\bet
(1-\al-\bet)^{\Delta-2}
\psi(\bet z_{1}+(1-\al)z_{2}, \al z_{2}+(1-\bet)z_{1})\ . \ret
&&
\label{conf}
\eeqar
Then from the definition of the
$ \hpsi$--function one has
\begin{equation}
 \hpsi_{h}(z_{1},z_{2})=\int_{0}^{\infty}dt_{1}dt_{2}t_{1}^{\Delta-1}
t_{2}^{\Delta-1}\psi_{h}(z_{1}t_{1},z_{2}t_{2})\rm{e}^{-t_{1}-t_{2}}.
\label{transf}
\end{equation}
After a change of variables
($ t_{1}+t_{2}=T $,
$ t_{1}=Ts $)  we get
\begin{eqnarray}
\lefteqn{\hpsi_{h}(z_{1},z_{2})=
\int_{0}^{\infty}dT\,
T^{2\Delta-1}\rm{e}^{-T}
\int_{0}^{1}d\gamma\int_{0}^{1}d\al\int_{0}^{{1-\al}}d\bet
(\gamma(1-\gamma))^{\Delta-1}} \nonumber \\
&& \nonumber \\
&&\times (1-\al-\bet)^{\Delta-2}
\psi(T(\bet\gamma z_{1}+(1-\al)(1-\gamma)z_{2}), T(\al(1-\gamma)
z_{2}+(1-\bet)\gamma z_{1})).
\label{new}
\end{eqnarray}
We perform the following change of variables in the integral:
\begin{equation}
{\bet\gamma=\nu s; \>\> (1-\gamma)(1-\al)=(1-s)\nu; \>\>
\al(1-\gamma)=t(1-\nu); \>\> (1-\bet)\gamma=(1-t)(1-\nu)}
\label{change}
\end{equation}
The last equation obviously follows from the three other equations.
Then it is easy to derive the following
relations:
\begin{equation}
0\leq \nu\leq 1;  \>\>\  0\leq s+t \leq 1
\end{equation}
and
\begin{equation}
 \gamma(1-\gamma)d\al d\bet d\gamma=\nu(1-\nu)d\nu ds dt\ ,
\end{equation}
\begin{equation}
 (1-\al-\bet)^{\Delta-2}(\gamma(1-\gamma))^{\Delta-2}=
 (1-s-t)^{\Delta-2}(\nu(1-\nu))^{\Delta-2}.
\label{self}
\end{equation}
Combining everything one immediately finds
\begin{equation}
\hpsi_{h}(z_{1},z_{2})=\int_{0}^{1}ds\int_{0}^{{1-s}}dt
(1-s-t)^{\Delta-2}\hpsi(sz_{1}+(1-s)z_{2}, tz_{2}+(1-t)z_{1})=
h_{12}\hpsi(z_{1},z_{2}).
\end{equation}
\vfill
\pagebreak

\pagebreak \thispagestyle{empty}
\setcounter{section}{1}

\noindent
\subsection*{Figure captions}

\renewcommand{\labelenumi}{{Fig.} \arabic{enumi}.}
\begin{enumerate}
\item Cancellation of the self--energy diagram contribution
of the $\sigma$--field.
\item Only diagram contributing in first order in~$1/N$
for operator insertions from ${\cal C}_{\rm sym}$.
\item The eigenvalues $\lambda$ obtained by diagonalizing
(\ref{eqhpsi}) in the subspace of conformal operators with
$n=3$~fields plotted as a function of~$d=2\mu$. The resp.~critical
exponents and anomalous dimensions can be obtained
via Eq.~(\ref{critdim}). The number of gradients~$l$ is
only given for the curves approaching the dashed limit
curve~$\phi^2\otimes\phi$, the unlabelled curves approach
other limit curves in agreement with our theorem. Notice
that the normalization factor~$2/\G(d/2-2)$ has been chosen
such that the values for $d=2$ agree with the eigenvalues
$\lambda_{2+\eps}$ in our definition~(\ref{twoplusx}),
same for $d=4$ and $\lambda_{4-\eps}$
using~(\ref{fourminusx}).
\item The function $K(d)$ from
(\ref{hg1NK}) plotted as a function of~$d$. Observe the sign
change in three dimensions separating stable from seemingly
unstable behaviour.
\end{enumerate}

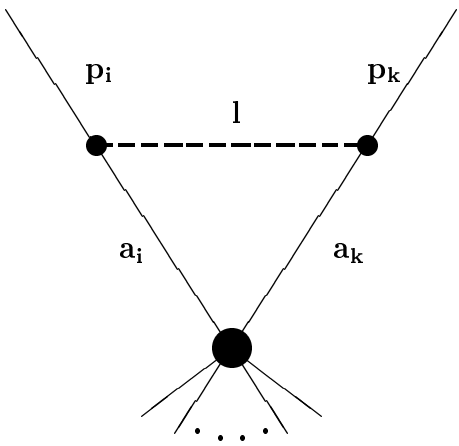
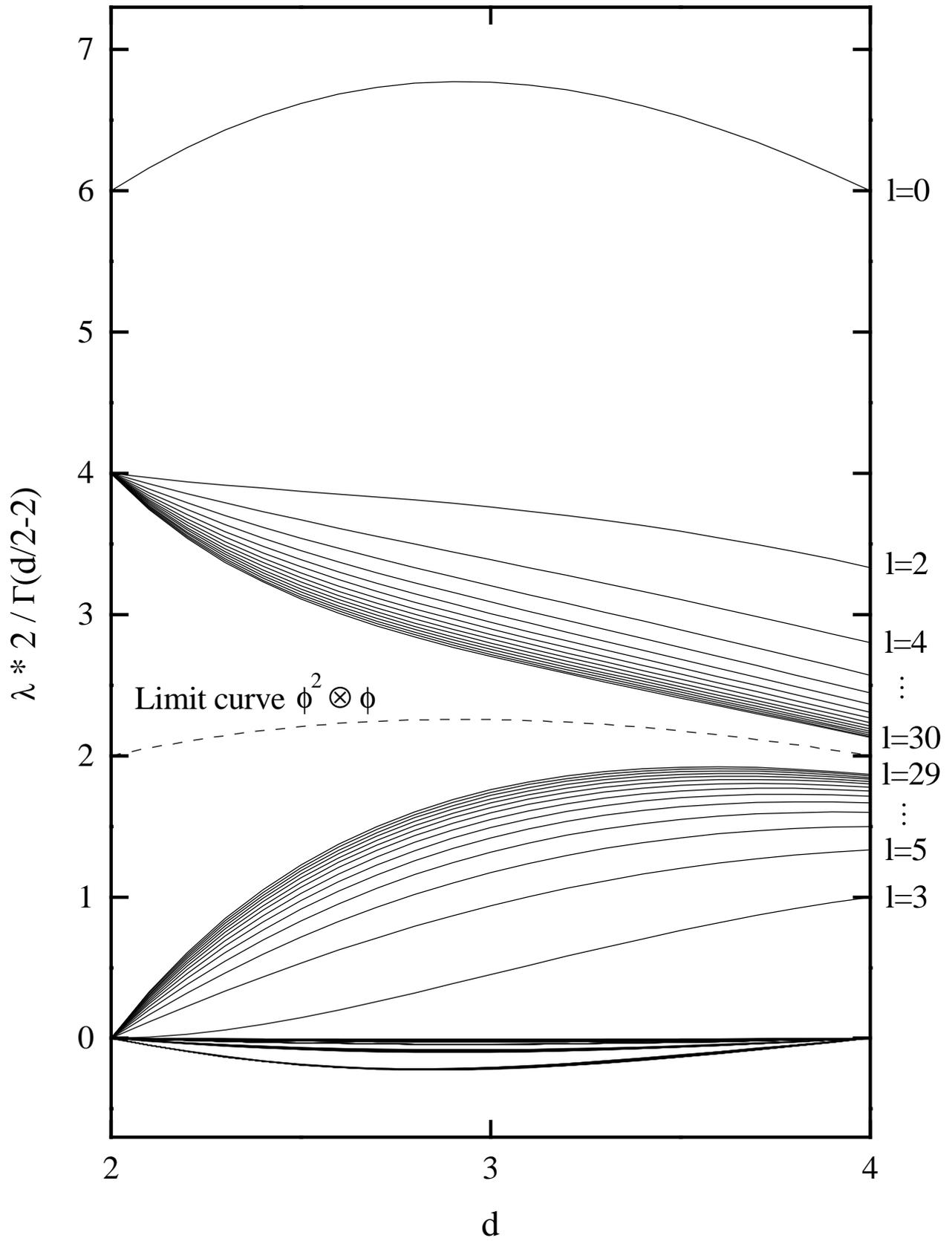
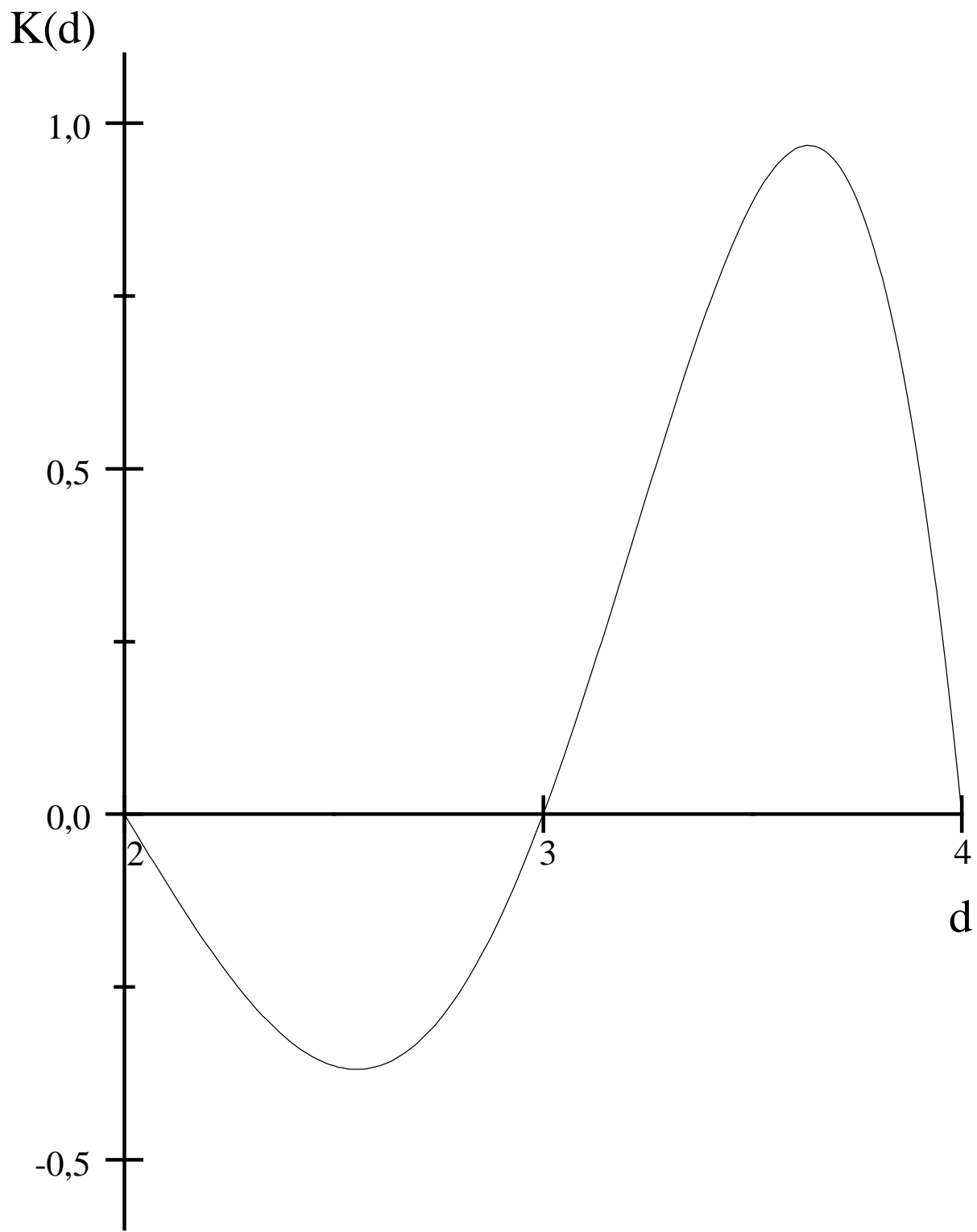
\end{document}